\documentstyle[12pt,psfig]{article}

\def\eea{\end{eqnarray}}
\def\bea{\begin{eqnarray}}
\def\eeas{\end{eqnarray*}}
\def\beas{\begin{eqnarray*}}
\def\ee{\end{equation}}
\def\be{\begin{equation}}

\def\tr{\mbox{Tr}\,}

\def\res{\mbox{Res}}
\def\re{\mbox{Re}\,}

\renewcommand{\thefootnote}{\fnsymbol{footnote}}
\begin{document}
\begin{titlepage}

\begin{center}

\vspace*{.5cm}
\hfill LA PLATA--TH 00/13

\vspace*{1cm}

{\large\bf CHIRAL PHASE TRANSITION IN A
COVARIANT NONLOCAL NJL MODEL} \vskip 1.5cm

{I. GENERAL$^a$, D. GOMEZ DUMM$^b\ ^\dagger$ and N. N. SCOCCOLA$^{a,c}$
\footnote[2]{Fellow of CONICET, Argentina.}} \vskip .2cm
{\it
$^a$ Physics Department, Comisi\'on Nacional de Energ\'{\i}a At\'omica,
      Av.Libertador 8250, (1429) Buenos Aires, Argentina.\\
$^b$ IFLP, Depto.\ de F\'{\i}sica, Universidad Nacional de La Plata, \\
C.C. 67, (1900) La Plata, Argentina.\\
$^c$ Universidad Favaloro, Sol{\'\i}s 453, (1078) Buenos Aires, Argentina. }

\vskip .5cm
October 2000 (revised March 2001)

\vskip .5cm {\bf ABSTRACT}\\
\begin{quotation}
The properties of the chiral phase transition at finite temperature
and chemical potential are investigated within a nonlocal covariant
extension of the Nambu-Jona-Lasinio model based on a separable
quark-quark interaction. We consider both the situation in which
the Minkowski quark propagator has poles at real energies and the
case where only complex poles appear. In the literature, the latter
has been proposed as a realization of confinement. In both cases,
the behaviour of the physical
quantities as functions of $T$ and $\mu$ is found to be quite
similar. In particular, for low values of $T$ the chiral
transition is  always of first order and, for finite quark masses,
at certain ``end point" the transition turns into a smooth
crossover. In the chiral limit, this ``end point" becomes a
``tricritical" point. Our predictions for the position of these
points are similar, although somewhat smaller, than previous
estimates. Finally, the relation between the deconfining
transition and  chiral restoration is also discussed.
\end{quotation}
\end{center}

\vskip 1.cm


\end{titlepage}

\renewcommand{\thefootnote}{\arabic{footnote}}
\setcounter{footnote}{0}

\vskip 1cm
The behaviour of hot dense hadronic matter and its transition to a plasma
of quarks and gluons has received considerable attention in recent years.
To a great extent this is motivated by the advent of facilities like e.g.\
RHIC at Brookhaven which are expected to provide some empirical information
about such transition. The interest in this topic has been further increased
by the recent suggestions that the QCD phase diagram could be richer than
previously expected (see Ref.\cite{Wil00} for some recent review articles).
Due to the well known difficulties to deal directly with
QCD, different models have been used to study this sort of problems. Among them
the Nambu-Jona-Lasinio model\cite{NJL61} is one of the most popular. In this
model the quark fields interact via local four point vertices which are subject to
chiral symmetry. If such interaction is strong enough the chiral symmetry is spontaneously
broken and pseudoscalar Goldstone bosons appear. It has been shown by many authors
that when the temperature and/or density increase, the chiral symmetry is
restored\cite{VW91}. For zero chemical potential and finite temperature this transition
is found to be a smooth one. However, whether for finite chemical potential and
zero temperature the transition is of first order or not is highly dependent on
the parameters of the model and on the approximations made. For example, in the $SU(2)$ version
of the NJL model a Hartree-Fock treatment leads to a first order transition, while the Hartree
treatment as well as the $SU(3)$ model calculations seem to favor a second order one\cite{VW91}.
On the other hand, several recent calculations performed within various
other models\cite{ARW98,BR98,HJSSV98} clearly
suggest that in QCD this transition should be of first order. Interestingly, the
model used in Ref.\cite{BR98} can be understood as a nonlocal generalization of the NJL.
However, the corresponding four point vertex is local in time and, therefore, not covariant.
Some covariant nonlocal extensions of the NJL model have been studied in the last few years\cite{Rip97}.
Nonlocality arises naturally in the context of several of the most successful approaches
to low-energy quark dynamics as, for example, the instanton liquid model\cite{SS98} and the
Schwinger-Dyson resummation techniques\cite{RW94}. It has been also argued that nonlocal
covariant extensions of the NJL model have several advantages over the local scheme.
Namely, nonlocal interactions regularize the model in such a way that anomalies are
preserved\cite{AS99} and charges properly quantized, the effective interaction is
finite to all orders in the loop expansion and therefore there is not need to introduce
extra cut-offs, soft regulators such as Gaussian functions lead to small NLO
corrections\cite{Rip00}, etc. In addition, it has been shown\cite{BB95} that a proper
choice of the nonlocal regulator and the model parameters can lead to some form of quark confinement, in
the sense of a quark propagator without poles at real energies. Recently, the behaviour
of this kind of models at finite temperature has been investigated\cite{BBKMT00}. In
this work we extend such studies to finite temperature and chemical potential.

We consider a nonlocal extension of the SU(2) NJL model defined by the effective action
\begin{eqnarray}
S &=& \int d^4x \ {\bar \psi}(x) \left( i \rlap/\partial  - m_c \right) {\psi}(x) +
       \int d^4x_1 ... d^4x_4 \ V(x_1,x_2,x_3,x_4)  \nonumber \\
  & & \qquad \qquad \times
  \left( {\bar \psi}(x_1) \psi(x_3) {\bar \psi}(x_2) \psi(x_4) +
         {\bar \psi}(x_1) i \gamma_5 \tau^a \psi(x_3) {\bar \psi}(x_2) i \gamma_5 \tau^a \psi(x_4)
         \right)\,,
\end{eqnarray}
where $m_c$ is the (small) current quark mass responsible for the explicit chiral
symmetry breaking. The interaction kernel in Euclidean momentum space is given by
\begin{equation}
V(q_1,q_2,q_3,q_4) = \frac{(2\pi)^4}{2} \ G \ r(q_1^2)  r(q_2^2) r(q_3^2) r(q_4^2)
\ \delta(q_1+q_2-q_3-q_4)\,,
\end{equation}
where $r(q^2)$ is a regulator normalized in such a way that $r(0) = 1$. Some general
forms for this regulator like Lorentzian or Gaussian functions have been used in the
literature. A particular form is given e.g.\ in the case of instanton liquid models.

Like in the local version of the NJL model, the chiral symmetry is spontaneously
broken in this nonlocal scheme for large enough values of the coupling $G$. In
the Hartree approximation the self-energy $\Sigma(q^2)$ at vanishing temperature
and chemical potential is given by
\begin{equation}
\Sigma(q^2) =  m_c + (\Sigma(0) - m_c) r^2(q^2)\,,
\end{equation}
where the zero-momentum self-energy $\Sigma(0)$ is a solution of the gap equation
\begin{equation}
\frac{2\pi^4}{G \ N_c} \left( \Sigma(0) - m_c \right) =
\int d^4q \ \frac{ \left[ m_c + (\Sigma(0) - m_c) r^2(q^2)\right] r^2(q^2)}
               {q^2 + \left[ m_c + (\Sigma(0) - m_c)
               r^2(q^2)\right]^2}\,.
\label{gapeq0}
\end{equation}
In general, the quark propagator might have a rather complicate structure of poles
and cuts in the complex plane.
In what follows we will assume that the regulator is such that it only has
an arbitrary but numerable set of poles. As already mentioned,
the absence of purely imaginary poles in the Euclidean quark propagator might
be interpreted as a realization of confinement\cite{BB95}. In that case quartets of
poles located at $\alpha_p = R_p \pm i \ I_p$, $\alpha_p = - R_p \pm i \ I_p$
appear. On the other hand, if purely imaginary poles exist they will show up as
doublets $\alpha_p = \pm i \ I_p$.  It is clear that the number and position of the
poles depend on the details
of the regulator. For example, if we assume it to be a step function as in the standard
NJL model only two purely imaginary poles at $\pm i \ M$ appear, with $M$ being the
dynamical quark mass. For a Gaussian interaction,
three different situations might occur. For values of $\Sigma(0)$ below a certain critical value
$\Sigma(0)_{crit}$ two pairs of purely imaginary simple poles and an infinite set of
quartets of complex simple poles appear. At $\Sigma(0) = \Sigma(0)_{crit}$, the two
pairs of purely imaginary simple poles turn into a doublet of double poles with
$R_p=0$, while for $\Sigma(0) > \Sigma(0)_{crit}$ only an infinite set of quartets of
complex simple poles is obtained. For the Lorentzian interactions there is
also a critical value above which purely imaginary poles cease to exist. However, for
this family of regulators the total number of poles is always finite.

To introduce finite temperature and chemical potential we follow the imaginary time formalism.
Thus, we replace the fourth component of the Euclidean quark momentum by $\omega_n - i \mu$,
where $\omega_n = (2 n + 1) \pi T$ are the discrete Matsubara frequencies and $\mu$ is
the chemical potential. In what follows we will assume that the temperature and chemical
potential dependences enter only through those quantities that carry a $q^2$-dependence
at $T=\mu=0$. That is to say, we will consider that the model parameters $G$ and $m_c$,
as well as the shape of the regulator, do not change with $T$ or $\mu$.
Performing this replacement in the gap equation Eq.(\ref{gapeq0}) we get
\begin{equation}
\frac{2\,\pi^4}{G \ N_c} \left( \Sigma(0) - m_c \right) = 2\pi\,T \int d^3\vec q
\ \sum_{n=-\infty}^{n=\infty} \ {\cal F}(q_{i \omega_n})\,,
\label{gapeq}
\end{equation}
where
\begin{equation}
{\cal F}( q ) = \frac{ \Sigma (q^2) r^2(q^2)}
               { q^2 + \Sigma^2(q^2)}
\end{equation}
and $q_z^2 = (- i z - i \mu)^2 + \vec q^{\ 2}$.  The sum over $n$ can
be expressed in terms of a sum over the poles of ${\cal F}( q_z )$, that we denote $z_p$, by
introducing the auxiliary function $f(z) = 1/(1+\exp(z/T))$ and using
the standard techniques described e.g.\ in Ref.\cite{K89}. It is easy to see
that the poles of ${\cal F}( q_z )$ are closely related to those of the quark
propagator $\alpha_p$. For the pair $\alpha_p = \pm (R_p + i I_p)$ the associated
values are $z_p = \pm \epsilon_p - \mu \mp i R_p I_p/\epsilon_p$, while for
$\alpha_p = \pm (R_p - i I_p)$ one has
$z_p = \pm \epsilon_p - \mu \pm i R_p I_p/\epsilon_p$, where $\epsilon_p$ is
defined as
\begin{equation}
\epsilon_p = \sqrt{ \frac{ I_p^2 - R_p^2 + \vec q\ ^2 +
\sqrt{  (I_p^2 - R_p^2 + \vec q\ ^2)^2 + 4 R_p^2 I_p^2 }}{2}}\;.
\end{equation}
Assuming that all the poles are
simple\footnote{This assumption is made just to keep our expressions into a simple
form. The generalization to poles of arbitrary multiplicity is rather
straightforward.} and performing explicitly the sum over all the poles in a given
multiplet we get
\begin{eqnarray}
T \int d^3\vec q \sum_{n=-\infty}^{n=\infty}
 \ {\cal F}(q_{i \omega_n}) &=& \frac{1}{2\pi}\,\int d^4 q \ {\cal F}(q) \nonumber \\
  -  \int d^3\vec q \ \sum_{\alpha_p} \ \!\! ' \ & & \!\!\!\!\!\!\!\!\!\!\!\!\!\!
\gamma_p \ \re\!\left[ \res[ 2z \ {\cal F}(z); \alpha_p]
\ \frac{\epsilon_p}{\epsilon_p^2 + i R_p I_p} \left( n_+ + n_- \right) \right]\,,
\label{sum}
\end{eqnarray}
where Res$[2 z{\cal F}(z); \alpha_p]$ stands for the residue
of the function $2 z {\cal F}(z)$ evaluated at $z=\alpha_p$,
and the prime in the sum indicates that it runs over all the poles
$\alpha_p = R_p + i I_p$ with $R_p \ge 0$ and $I_p>0$. The coefficient
$\gamma_p$ is defined as $\gamma_p=1/2$ for $R_p=0$ and $\gamma_p=1$ otherwise.
The generalized occupation numbers $n_\pm$ are given by
\begin{equation}
n_\pm = \left[ 1+ \exp\left( \frac{\epsilon_p  \mp \mu +
i R_p I_p/\epsilon_p}{T} \right) \right]^{-1}\,.
\end{equation}
As customary, in writing Eq.(\ref{sum}) we have isolated a term which has the same form
as the $T=\mu=0$ expression. In this way, all the $T$ and $\mu$ dependent contributions
remain finite. Replacing Eq.(\ref{sum}) in the right hand
side of Eq.(\ref{gapeq}) and using
\begin{equation}
\res[ 2 z \ {\cal F}(z); \alpha_p] = \left.
 \frac{\Sigma(u) r^2(u)}
{1+ \partial_{u} \Sigma^2(u) }\right|_{u=\alpha_p^2}
\end{equation}
we obtain the final form of the gap equation
at finite temperature and chemical potential. As we see the dependence on these
quantities is completely fixed by the pole structure of the quark propagator.

The other quantities which are of interest to understand the characteristics of
the chiral phase transition are the quark condensate $\langle\bar q q\rangle$
and the quark density $\langle q^\dagger q\rangle$ for each flavour.
At $T=\mu=0$, the condensate
is given by
\begin{equation}
\langle\bar q q\rangle = - \int \frac{d^4 q}{(2\pi)^4}
\ \tr \left[ \left( \rlap/q + \Sigma(q^2) \right)^{-1} \right]\,.
\end{equation}
Following similar steps as before, the corresponding result for
finite $T$ and $\mu$ can be cast into the form
\begin{eqnarray}
\langle\bar q q \rangle &=& - \frac{N_c}{4 \pi^4} \int d^4q\ \frac{\Sigma(q^2)}
 {q^2 + \Sigma^2(q^2)} \nonumber \\
& &  + \frac{N_c}{2\pi^3} \int d^3\vec q \
 \sum_{\alpha_p} \ \!\! ' \
 \gamma_p\ \re\left[ \left.
 \frac{\Sigma(u)}{1+ \partial_{u} \Sigma^2(u) }\right|_{u=\alpha_p^2}
\ \frac{\epsilon_p}{\epsilon_p^2 + i R_p I_p} \left( n_+ + n_- \right)\right] .
\label{qbq}
\end{eqnarray}
Away from the chiral limit this expression turns out to be divergent.
Following the standard procedure, we regularize it
by subtracting the value obtained in the absence of interactions.

To determine the quark density one has to be more careful due to the presence of
nonlocal interactions. In fact, they imply the existence of extra
contributions to the Noether currents. The proper expression for the quark
density in the Hartree approximation reads
\begin{equation}
\langle q^\dagger q\rangle = -i \int \frac{d^4 q}{(2\pi)^4} \, \tr
\left[ \left( \rlap/q + \Sigma(q^2) \right)^{-1}
\left(\gamma_4 + \partial_{q_4} \Sigma(q^2) \right) \right]\,.
\label{qq}
\end{equation}
It is seen here that, for any quark flavour,
the residue of the pole given by the dressed quark propagator is equal
to one. As shown in Ref.\cite{GBR98}, this leads to the correct normalization
of the baryon number, independently of the shape of the regulator.
As above, the result in Eq.(\ref{qq}) can be now extended to finite temperature
and chemical potential by replacing $q_4\to \omega_n-i\,\mu$ and summing over all
Matsubara frequencies $\omega_n$. In this case we obtain
\begin{equation}
\langle q^\dagger q\rangle = \frac{N_c}{2\pi^3} \int d^3\vec q \
 \sum_{\alpha_p} \ \!\! ' \ \gamma_p \ \re \left[  n_+ - n_- \right] .
\label{qdq}
\end{equation}

Having introduced the formalism needed to extend the model to finite temperature
and chemical potential we turn now to our numerical calculations. In this work we
take the nonlocal regulator to be of the Gaussian form
\begin{equation}
r(q^2) = \exp\left( - \frac{q^2}{2\Lambda^2} \right)
\label{ff}
\end{equation}
and consider two sets of values for the parameters of the model. Set I corresponds to
$G = 50$ GeV$^{-2}$, $m_c = 10.5$ MeV and $\Lambda = 627$ MeV, while for Set II the
respective values are $G = 30$ GeV$^{-2}$, $m_c = 7.7$ MeV and $\Lambda = 760$ MeV.
Both sets of parameters lead to the physical values of the pion mass and decay
constant. For Set I the calculated value of the chiral quark condensate at zero
temperature and chemical potential is $- (200$ MeV$)^3$ while for Set II it
is $- (220$ MeV$)^3$. These values are similar in size to those determined from
lattice gauge theory or QCD sum rules. The corresponding results for the self-energy
at zero momentum are $\Sigma (0) = 350$ MeV for Set I and $\Sigma (0) = 300$ MeV for
Set II. Using the explicit expression of $\Sigma(0)_{crit}$ for the Gaussian
interaction,
\begin{equation}
\Sigma(0)_{crit} = m_c + \frac{1}{2} \left( \sqrt{m_c^2 + 2 \Lambda^2} - m_c \right)
\exp\left[ - \frac{  \left( \sqrt{m_c^2 + 2 \Lambda^2}
+ m_c \right)^2}{4\,\Lambda^2}\right]\,,
\label{crit}
\end{equation}
it is easy to check that Set I corresponds to a situation in which there are no purely
imaginary poles of the Euclidean quark propagator and Set II to the case in which
there are two pairs of them. Therefore, we will also refer to Set I as the
confining set and to Set II as the non-confining one.

The behaviour of the zero-momentum self-energy $\Sigma (0)$, the chiral quark condensate
$\langle\bar q q\rangle$ and the quark density $\rho = \langle q^\dagger q\rangle$ as
functions of the chemical potential for some values of the temperature is shown in Fig.\ 1.
The left and right panels in the figure correspond to the results for Set I and
Set II, respectively. In both cases we observe the existence of some kind of
phase transition at (or around) a given value of the chemical potential which depends
on the temperature. To obtain our results, we have included in the sums
over $\alpha_p$ appearing in Eqs.(\ref{sum}), (\ref{qbq}) and (\ref{qdq}) the first
few poles of the quark propagator. We have checked, however, that for the range of
values of $T$ and $\mu$ covered in our calculations, the convergence is so fast that
already the first pole gives almost 100\% of the full result. Thus, the behaviour
of relevant physical quantities up to (and somewhat above) the phase
transition is basically dominated by the first pole of the quark propagator.

We observe in Fig.\ 1 that at $T=0$ there is a first order phase transition for both
the confining and the non-confining sets of parameters.
As the temperature increases, the
value of the chemical potential at which the transition shows up decreases. Finally,
above a certain value of the temperature the first order phase transition does not
longer exist and, instead, there is a smooth crossover. This phenomenon is clearly shown
in the right panel of Fig.\ 2, where we display the critical temperature at which the
phase transition occurs as a function of the chemical potential. The point at which
the first order phase transition ceases to exist is usually called ``end point".
In the chiral limit the latter turns into the so-called ``tricritical
point", which is the point at which the second order phase transition expected to
happen in QCD with two massless quarks becomes a first order one.
In fact, this is also what happens within the present model in the chiral limit,
as it is shown in the left panel of Fig.\ 2. Some predictions\cite{BR98,HJSSV98}
about both the position of this point and its possible
experimental signatures\cite{SRS99} exist in the literature. In our case the
``tricritical point" is located at $(T_P,\mu_P) = $(70 MeV$,\,130$ MeV) for
Set I and (70 MeV$,\,140$ MeV) for Set II, while the ``end points" are placed
at $(T_E,\mu_E) = $(70 MeV$,\,180$ MeV) and (55 MeV$,\,210$ MeV), respectively.
As we see the predicted values are very similar for both sets of parameters and
slightly smaller than the values in Refs.\cite{BR98,HJSSV98},
$T_P \approx 100$ MeV and $\mu_P \approx 200 - 230$ MeV. In this sense, we
should remark that our model predicts a critical temperature at $\mu = 0$ of
about 100 MeV, somewhat below the values obtained in modern lattice
simulations\cite{Kha00} which suggest $T_c \approx 140 - 190$ MeV. In any case,
our calculation seems to indicate that $\mu_P$ might be smaller than previously
expected even in the absence of strangeness degrees of freedom.

It is interesting to discuss in detail the situation concerning the confining set.
In this case we can find, for each temperature, the chemical potential
$\mu_{d}$ at which confinement is lost. Following the proposal of Ref.\cite{BB95},
this corresponds to
the point for which the self-energy at zero momentum reaches $\Sigma(0)_{crit}$
(c.f. Eq.(\ref{crit})). Using the values of $m_c$ and $\Lambda$ corresponding to
Set I we get $\Sigma(0)_{crit} = 267$ MeV.  For low temperatures,
$\mu_{d}$ coincides with the the chemical potential at which the
chiral phase transition takes place. However, for a temperature close enough to
that of the ``end point", $\mu_{d}$ starts to be slightly smaller than the value of
$\mu$ that corresponds to the chiral restoration. Above $T_E$ it is difficult to make
an accurate comparison since, for finite quark masses, the chiral restoration
proceeds through a smooth crossover. However, we can still study the situation in
the chiral limit. In this case we find that, in the region where the chiral
transition is of second order, deconfinement always occurs, for fixed $T$,
at a lower value of $\mu$ than the chiral transition. The corresponding
critical line is indicated by a dashed line in the left panel of Fig.\ 2. In
any case, as we can see in this figure, the departure of the line of chiral
restoration from that of deconfinement is in general not too large. This
indicates that within the present model both transitions tend to happen at,
approximately, the same point.

In conclusion, in this work we have investigated a nonlocal covariant
extension of the NJL model at finite temperature and chemical potential. We have
assumed that the corresponding four point vertex is separable in momentum space and
that the regulator leads to an Euclidean quark propagator with an arbitrary (but
discrete) number of poles in the complex plane.
We have shown that the dependence on $\mu$ and $T$ can be expressed in terms of
these poles and the corresponding residues. We have studied in detail the
particular case of a Gaussian regulator, considering two different situations.
In the first case, the Euclidean quark propagator does not have any
purely imaginary poles, which can be understood as a way of confinement.
In the second case, this type of poles exists, thus quarks would be not confined.
We have found that in both cases the behaviour of the physical quantities as
a function of $T$ and $\mu$ is quite similar. In particular, for low values
of $T$ the chiral transition
is always of first order and, for finite quark masses, at certain ``end point" the
transition turns into a smooth crossover. In the chiral limit, this ``end point"
becomes a ``tricritical" point. Our predictions for the position of these
points are similar, although somewhat smaller, than previous estimates obtained
in alternative models\cite{BR98,HJSSV98}. It is clear that several extensions of
this work are possible. Firstly, it would be interesting to include
interactions in the vector and axial-vector meson channels, in order to
investigate meson properties close to the phase
transition. Secondly, to study the possible
existence of color superconductivity in this sort of models, interactions
in the quark-quark channels should be incorporated. Finally, the extension
to the $SU(3)$ sector would allow to understand the role of the
strangeness in the characteristics of the chiral phase transition
as described by nonlocal covariant models. We hope to report
on these issues in forthcoming publications.

\vspace{1cm}

NNS wishes to thank O.\ Civitarese, H.\ Forkel and C.L.\ Schat for
very enlightening discussions. The authors are also thankful to
D.\ Blaschke for a very timely and useful correspondence. This
work was supported in part by a grant from Fundaci\'on Antorchas,
Argentina, and the grant PICT 03-00000-00133 from ANPCYT, Argentina.

\begin{figure}
\centerline{\psfig{figure=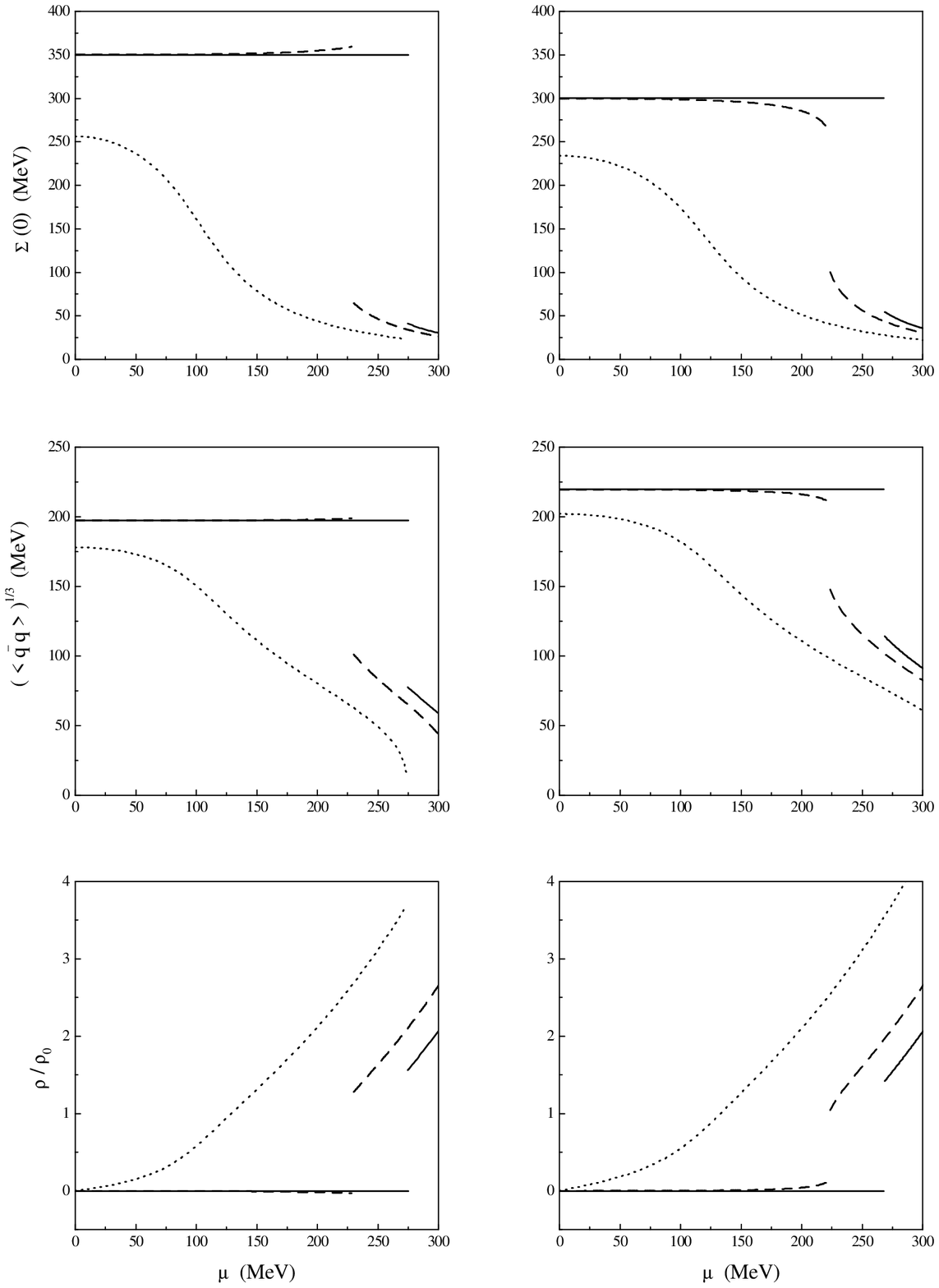,height=19cm}}
\protect\caption{\it Behaviour of the self-energy, the chiral condensate and
the quark density as functions of the chemical potential for three
representative values of the temperature. Full line corresponds to T=0, dashed
line to $T=50$ {\rm MeV} and dotted line to $T=100$ {\rm MeV}. The left panels
display the results for Set I and the right panels those for Set II.
The quark density $\rho$ is given with respect to nuclear matter density
$\rho_0 = 1.3\times 10^6$ {\rm MeV}$^3$.}
\label{tmdep}
\end{figure}

\pagebreak

\begin{figure}
\centerline{\psfig{figure=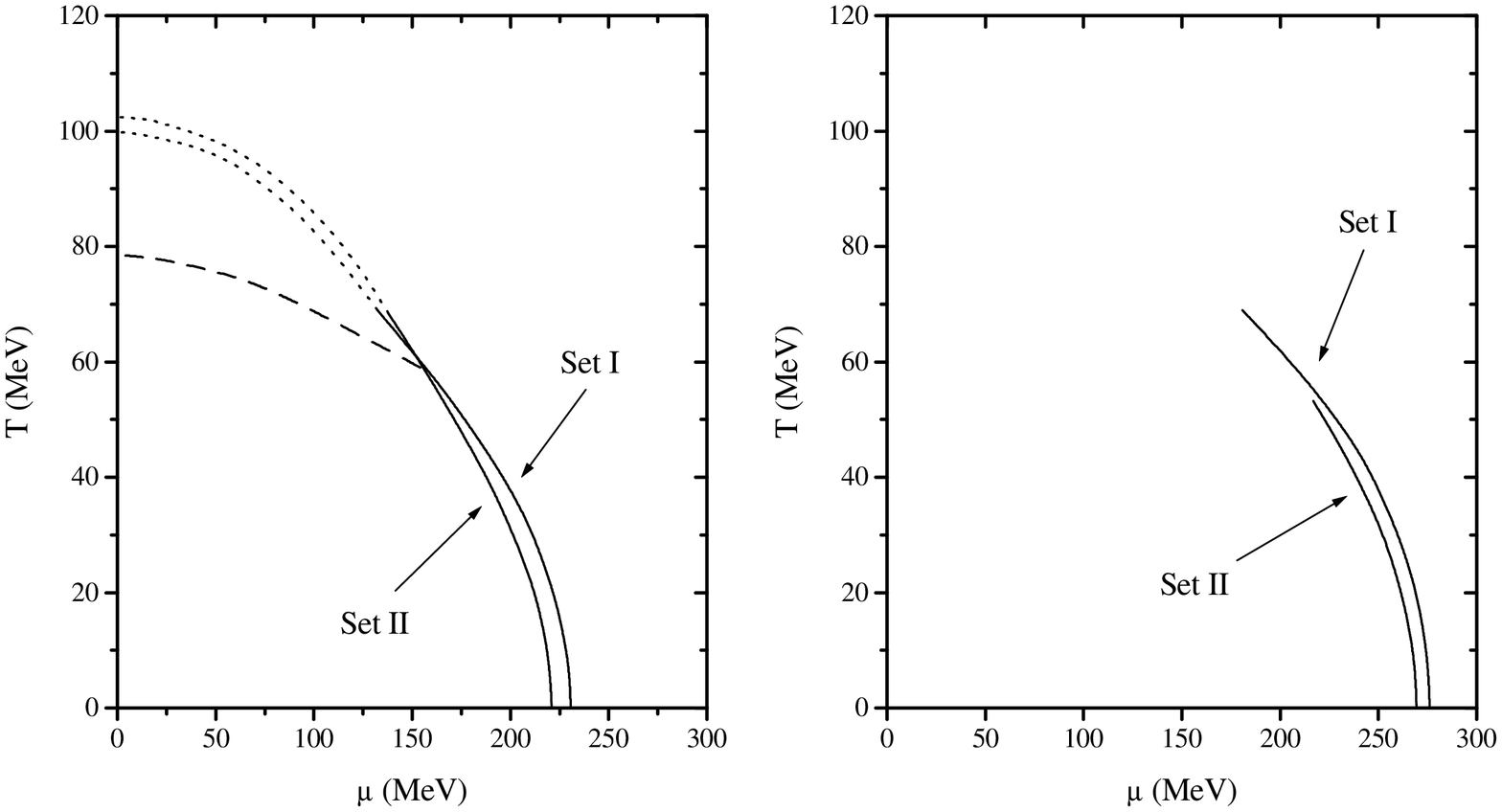,height=8cm}}
\protect\caption{\it Critical temperatures as a function of the chemical potential.
The left panel corresponds to the chiral limit and the right panel
to the case of finite quark masses. In the left panel, the dotted lines correspond
to the region of the second order phase transition, while the full lines in both
panels indicate the region where a first order transition occurs. The dashed line
in the left panel indicates the critical line for deconfinement corresponding to
Set I. For chemical potentials somewhat larger than $\mu_P$ this line coincides
with that of the chiral restoration.}
\label{trans}
\end{figure}


\begin{thebibliography}{99}
\bibitem{Wil00}
F.\ Wilczek,
{\it hep-ph/0003183};
K.\ Rajagopal,
{\it hep-ph/0009058}.

\bibitem{NJL61}
Y.\ Nambu and G.\ Jona-Lasinio,
Phys.\ Rev.\ {\bf 122} (1961) 345;
Phys.\ Rev.\ {\bf 124} (1961) 246.

\bibitem{VW91}
U.\ Vogl and W.\ Weise,
Prog.\ Part.\ Nucl.\ Phys.\ {\bf 27} (1991) 195;
S.\ Klevansky,
Rev.\ Mod.\ Phys.\ {\bf 64} (1992) 649;
T.\ Hatsuda and T.\ Kunihiro,
Phys.\ Rep.\ {\bf 247}, (1994) 221.

\bibitem{ARW98}
M.\ Aldorf, K.\ Rajagopal and F.\ Wilczek,
Phys.\ Lett.\ {\bf B422} (1998) 247, Nucl.\ Phys.\ {\bf B537} (1999) 443;
R.\ Rapp, T.\ Schafer, E.V.\ Shuryak and M.\ Velkovsky,
Phys.\ Rev.\ Lett.\ {\bf 81} (1998) 53.

\bibitem{BR98}
J.\ Berges and K.\ Rajagopal,
Nucl.\ Phys.\ {\bf B538} (1999) 215.

\bibitem{HJSSV98}
M.A.\ Halasz, A.D.\ Jackson, R.E.\ Shrock, M.A.\ Stephanov and J.J.M.\ Verbaarschot,
Phys.\ Rev.\ {\bf D58} (1998) 096007.

\bibitem{Rip97}
G.\ Ripka,
{\it Quarks bound by chiral fields} (Oxford University Press, Oxford, 1997).

\bibitem{SS98}
T.\ Schaefer and E.\ Schuryak,
Rev.\ Mod.\ Phys.\ {\bf 70} (1998) 323.

\bibitem{RW94}
C.D.\ Roberts and A.G.\ Williams,
Prog.\ Part.\ Nucl.\ Phys.\ {\bf 33} (1994) 477;
C.D.\ Roberts and S.M.\ Schmidt,
Prog.\ Part.\ Nucl.\ Phys.\ {\bf 45S1} (2000) 1.

\bibitem{AS99}
E.R.\ Arriola and L.L.\ Salcedo,
Phys.\ Lett.\ {\bf B450} (1999) 225.

\bibitem{Rip00}
G.\ Ripka,
{\it hep-ph/0003201};
R.S.\ Plant and M.C.\ Birse,
{\it hep-ph/0007340}.

\bibitem{BB95}
R.D.\ Bowler and M.C.\ Birse,
Nucl.\ Phys.\ {\bf A582} (1995) 655;
R.S.\ Plant and M.C.\ Birse,
Nucl.\ Phys.\ {\bf A628} (1998) 607.

\bibitem{BBKMT00}
D.\ Blaschke, G.\ Burau, Y.L.\ Kalinovsky, P.\ Maris and P.C.\ Tandy,
{\it nucl-th/0002024};
B.\ Szczerbi\'nska and W.\ Broniowski,
Acta Phys.\ Polon.\ {\bf B31} (2000) 835.

\bibitem{K89}
J.I.\ Kapusta, {\it Finite-temperature field theory}
(Cambridge Univ. Press, Cambridge, 1989).

\bibitem{GBR98}
B.\ Golli, W.\ Broniowski and G.\ Ripka,
Phys.\ Lett.\ {\bf B437} (1998) 24.

\bibitem{SRS99}
M.\ Stephanov, K.\ Rajagopal and E.\ Shuryak,
Phys.\ Rev.\ Lett.\  {\bf 81} (1998) 4816;
Phys.\ Rev.\ {\bf D60} (1999) 114028.

\bibitem{Kha00}
F.\ Karsch,
Nucl.\ Phys.\ (Proc.\ Suppl.) {\bf 83} (2000) 14;
A.\ Ali Khan {\it et al.},
Phys.\ Rev.\ {\bf D63} (2001) 034502.

\end{thebibliography}
\end{document}